# Ionic Bonds Control Ferroelectric Behavior in Wurtzite Nitrides

*Keisuke Yazawa[1,2], John Mangum[1], Prashun Gorai[1,2], Geoff L. Brennecka[2\*], and Andriy Zakutayev[1\*]*

1. Materials Science Center, National Renewable Energy Laboratory, Golden, Colorado 80401, United States
2. Department of Metallurgical and Materials Engineering, Colorado School of Mines, Golden, Colorado 80401, United States

**Corresponding Authors**

\*E-mail: Andriy.Zakutayev@nrel.gov and geoff.brennecka@mines.edu

**ABSTRACT**

Ferroelectricity enables key integrated technologies from non-volatile memory to precision ultrasound. Wurtzite ferroelectric $Al_{1-x}Sc_xN$ has recently attracted attention because of its robust ferroelectricity and Si process compatibility in addition to being the first known ferroelectric wurtzite. However, the origin and control of ferroelectricity in wurtzite materials is not yet fully understood. Here we show that the local bond ionicity, rather than simply the change in tetrahedral distortion, is key to controlling the macroscopic ferroelectric response, according to our coupled experimental and computational results. Across the composition gradient in Sc < 0.35 range and 140 – 260 nm thickness in combinatorial thin films of $Al_{1-x}Sc_xN$, the pure wurtzite phase exhibits a similar *c/a* ratio regardless of the Sc content, due to elastic interaction with neighboring crystals. The coercive field and spontaneous polarization significantly decrease with increasing Sc content despite this invariant *c/a* ratio, due to the more ionic bonding nature of Sc-N relative to the more covalent Al-N bonds, supported by DFT calculations. Based on these insights, ionicity engineering is introduced as an approach to reduce coercive field of $Al_{1-x}Sc_xN$ for memory and other applications and to control ferroelectric properties in other wurtzites.

**Main**

Ferroelectricity, which is an ability of a material to reverse spontaneous polarization of by applying electric field, was discovered more than a hundred years ago.[1] Since that time ferroelectric materials have enabled and advanced ubiquitous electronic technologies in the forms of capacitors, piezoelectric actuators/sensors, energy harvesters, non-linear optics, pyroelectric sensors, and PTC thermistors.[2–7] Thin film ferroelectrics have further enabled integrated devices such as FeRAM and PiezoMEMS.[8,9] Research continues into use of ferroelectrics for multiferroic[10] and photovoltaic[11,12] properties as well as and negative capacitance,[13] and new concepts keep emerging.[13–18]

The recently-developed wurtzite aluminum scandium nitride alloy ($Al_{1-x}Sc_xN$) ferroelectric has received significant attention because of its enhanced piezoelectric response,[19] robust ferroelectricity,[20] and compatibility with both Si and III-N semiconductor processes. Chemistry, stress, strain, and film thickness have been rigorously investigated to control ferroelectricity in this material system.[20–26] Higher Sc/Al ratio simultaneously reduces the crystallographic *c/a* ratio, spontaneous polarization, and coercive field,[20,27] and in-plane tensile strain lowers the coercive field.[20,23] Understanding how properties—particularly coercive field (voltage)—scale with film thickness is also important from a future device perspective.[25,26]

The prior studies of ferroelectricity in $Al_{1-x}Sc_xN$[20–26] included explicit and implicit variables such as residual strain state, process parameters, target condition, chamber type, substrate type/treatment, etc. These variables convolute experimental effects, and the resulting data scatter can easily mask important but unrepresented factors such as microstructure and defects. Combinatorial techniques reduce uncontrolled process variables because a single film library can include all samples of interest and also offer significant advantages for rapid screening.[28,29] In such high-throughput methods, experimental variables (e.g., composition, thickness, substrate temperature) are smoothly varied across a single sample library,

Previously, combinatorial ferroelectric property screening has been reported for perovskite materials based on $BaTiO_3$,[30,31] $Pb(Zr,Ti)O_3$,[32] $BiFeO_3$,[33,34] and $Bi_4Ti_3O_{12}$.[35] Those studies validated the combinatorial method for ferroelectric materials, but they missed the opportunity to directly address the effect of residual elastic strain in the sample library.[36] Lacking information about the underlying thermodynamic coefficients, it is difficult to separate elastic strain from total strain in order to isolate effects of electromechanical coupling and chemistry on ferroelectric properties in wurtzite ferroelectrics. Thus, systematic crystallographic analysis along with corresponding ferroelectric property evaluation in a combinatorial study is key for understanding the chemical and elastic contributions to ferroelectric response.

In this study, we decouple effects of composition and wurtzite *c/a* ratio on ferroelectric properties in $Al_{1-x}Sc_xN$ by investigating chemistry–structure–thickness–property relations using combinatorial libraries with simultaneous gradients in both composition and thickness. The pure wurtzite phase in the Al-rich region (Sc < 0.35) of the library exhibits a constant wurtzite *c/a* ratio regardless of Sc content likely due to elastic clamping across the composition gradient in the combinatorial library. In this composition region without measurable change in the crystal lattice, the remanent polarization and coercive field decrease with Sc content, similar to homogeneous films for which *c/a* ratio varies directly with composition. The composition-dependent and structure-independent change in ferroelectric properties is attributed to the more ionic nature of the Sc-N compared to Al-N bonds, as confirmed with DFT calculations. These results indicate that the nature of chemical bonding is more significant than crystallographic *c/a* ratio for determining ferroelectric properties in the wurtzite ferroelectrics and suggest new ways for future material discovery and development of ferroelectric wurtzite materials.

Analysis of the crystalline structure of the deposited $Al_{1-x}Sc_xN/Pt/TiO_x/SiO_2/Si$ library (Figure 1) shows textured pure wurtzite at relatively low Sc contents (x < 0.35), and mixed-phase wurtzite plus rocksalt at higher Sc content (x > 0.35). Two-dimensional ($2\theta$ - chi) diffraction mapping in a specific location (Sc content: 0.4, thickness: 230 nm) is a representative result of phase coexistence of the well-oriented wurtzite 002 and rocksalt 111 diffraction peaks along with Pt 111 (textured bottom electrode) and Au 111/200 peaks (polycrystalline top electrode) as shown in Figure 1(a). Diffraction profiles, each of which integrates a chi range of 72 degrees, taken from a series of points along the x-axis of the sample library (Sc content ranges from 0.25 to 0.50 while film thickness is similar, 230 – 260 nm) are stacked in Figure 1(b). The intensity of the wurtzite 002 peak is highest when Sc content is lowest. A shoulder peak at lower $2\theta$ angle attributed to the rocksalt phase appears when the Sc content is > 0.35. This wurtzite and rocksalt phase co-existence has also been seen in homogeneous films[19,37,38] and is associated with the wurtzite – rocksalt miscibility gap.[39] Figure 1(c) shows lattice parameters *a* and *c* for the pure wurtzite region in the thickness – composition space. The color scale shows the *c* lattice parameter, and the marker size represents the *a* lattice parameter. The *a* lattice parameter is calculated with the off axis wurtzite 103 diffraction peak shown in Figure S1 in Supporting Information. The *c* lattice parameter gradually increases up to Sc = 0.3 and then decreases, while the *a* lattice parameter keeps increasing up to Sc = 0.35. This trend is in good agreement with reported homogeneous films and calculations.[19,40]

Figure 1(d) represents the *c/a* ratio in the thickness – composition space. The value of *c/a* increases slightly from 1.58 to 1.59 with increasing Sc content towards Sc = 0.3 then decreases to 1.57 at Sc = 0.35. The change that we see in the *c/a* ratio across this sample library is notably less than the reported *c/a* reduction from 1.57 (Sc = 0.25) to 1.51 (Sc = 0.35) in homogeneous films, each grown separately,[19] This difference likely comes from the difference in elastic strain state in the gradient library, namely our film undergoes more out-of-plane tensile and/or in-plane compressive strains compared to the homogeneous film set, although the absolute elastic strains in both cases are unknown. This relative elastic strain likely originates from the composition gradient of the combinatorial deposition, i.e., because of the continuous composition gradient, the Sc-rich regions are elastically clamped by their stiffer Al-rich neighboring grains.[41] As a result, the library has a unique feature that the crystal lattice—and in particular, the *c/a* ratio—is essentially invariant with composition across the pure wurtzite region. There is also no significant thickness dependence on *c/a* ratio so that the thickness effect on the crystal lattice is minimal in the 140 – 260 nm range.

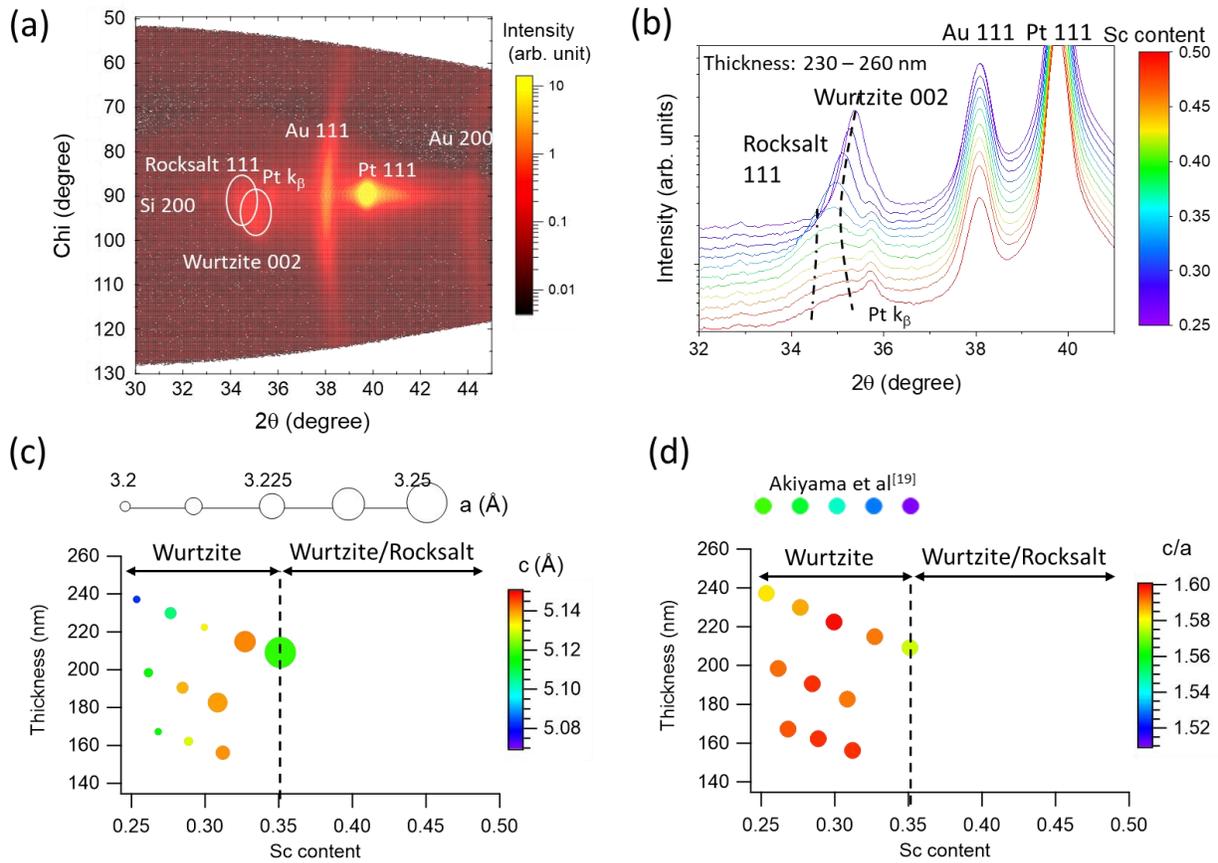

Figure 1. Crystal structure analysis using XRD. (a) 2D (2θ-chi) diffraction scan for Sc = 0.4 and thickness = 230 nm showing wurtzite/rocksalt coexistence. Chi 90 degree represents the surface normal direction (see Figure S2 in Supporting Information for the definition of the chi direction.) (b) Wurtzite peak shift in integrated 2θ profiles for varying Sc content from 0.25 to 0.5. (c) Lattice parameter map in composition – thickness space. (d) *c/a* ratio map in composition – thickness space. Comparison to homogeneous film data reported by Akiyama[19] indicates significantly invariant c/a ratio with regard to composition in this study.

Film microstructure, such as texture and grain orientation, influences ferroelectric properties and is therefore investigated in this work.[42–46] In combinatorial sputtering, the particle flux has a asymmetric directionality with respect to the substrate, which causes oblique grain growth in addition to the desired composition gradient(s). Cross-sectional TEM micrographs are shown in Figure 2(a) for films for Sc = 0.25, Sc = 0.38 and Sc = 0.50. The right sides of the cross-sectional micrographs correspond to the side nearest the Al target. Grain growth directs to the Al target, clearly seen in the Sc rich (far from the Al target) region, where grain boundary angles are ~10-15° from the surface normal (Sc = 0.50). The oblique angle is less (6 – 13°) across the middle portion (Sc = 0.38), and grain boundaries are nearly straight for the portion of the film directly above the Al target (Sc = 0.25). These results suggest that the Al flux is controlling the growth of the columnar grains.

Crystal misorientation is particularly important for ferroelectric properties in wurtzites because the spontaneous polarization is aligned with—and limited to—the *c*-axis of the wurtzite crystal. Chi–omega scans of the wurtzite 002 peak show the crystal misorientation (Figure 2(b)) at a sample position where Sc content is 0.47 and thickness is 140 nm. The 0 degree position in both chi and omega offsets aligns to the substrate Pt (111) peak. The wurtzite peak is -2 degrees off-centered in chi and -1 degrees off-centered in omega. Based on the chi and omega directions found in supplementary Figure S2 (Supporting Information), the (002) plane normal is directed slightly towards the Al target in both chi and omega, which is consistent with the grain growth direction seen in Figure 2(a). Figure 2(c) maps the chi and omega offsets with respect to the Pt (111) normal direction across the sample library. The color bar represents the omega offset, and the marker size shows the chi offset. The region close to the Al target (x = 44 and y = 6 mm) aligns well to Pt (111) in both omega and chi, and the offset angles in both chi and omega are larger at positions further away from the Al target, which supports Al flux-dominated growth. However, the angle is not as significant as the oblique grain growth angle shown in Figure 2(a). At most, the chi offset is -6 degrees, and omega offset is -3 degrees. This means that the effect of the flux directionality on crystal orientation exists but is limited compared to that on the grain growth direction. A 5° crystallographic misalignment corresponds to a 0.5% loss in remanent polarization because of the cosine projection; this is less than the uncertainties in either thickness or area, so the crystallographic misorientation is considered to be insignificant in this work.

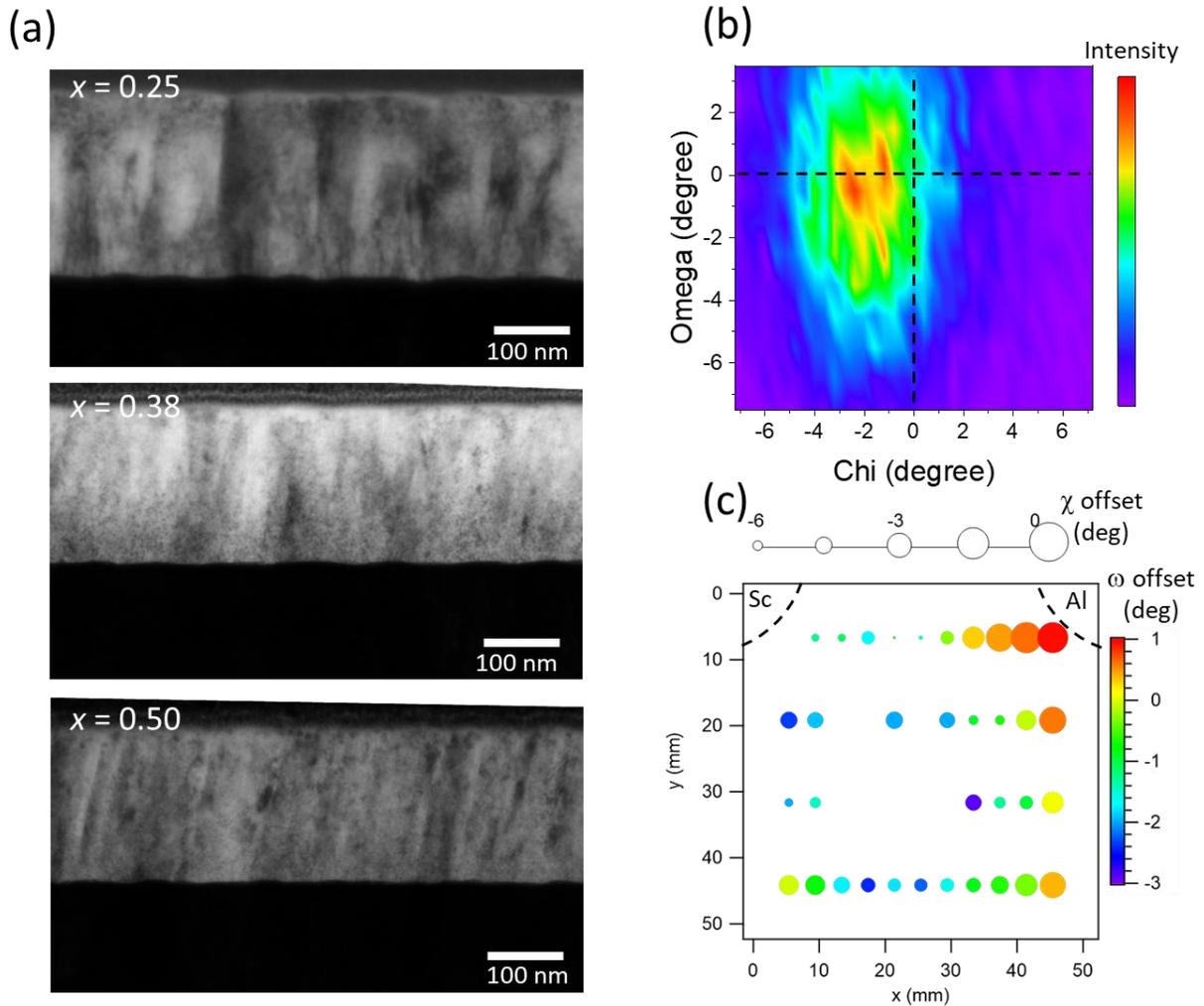

Figure 2. Grain and crystal growth direction analysis. (a) Cross-sectional TEM micrographs for Sc = 0.25 (under Al target), Sc = 0.38 (middle of library), and Sc = 0.5 (under Sc target) showing oblique grain growth. (b) XRD omega–chi plan view map for (002) wurtzite peak with respect to substrate Pt (111) normal direction. (c) Omega and chi offsets across the sample library mapping out crystal growth direction with respect to Pt (111) normal.

Electrical properties for the library are shown in Figure 3. Leakage current was measured using a bent probe with gauge of ~300 μm contacting the bare surface of the film. Compositional dependence (similar thickness ranging from 230 – 260 nm) of I-V measurements is shown in Figure 3(a). An exponential current increase is observed for Sc > 0.35, and the threshold voltage for the exponential increase decreases with increased Sc content. The threshold voltage is defined here as the voltage at which the current exceeds $4 \times 10^{-9}$ A (the dashed line in Figure 3(a)). Because the compositions of Sc ≥ 0.35 correspond to the presence of mixed wurtzite and rocksalt phases, the current increase is attributed to the presence of a conductive (bandgap of ~1eV) rocksalt phase.[47,48] The threshold voltage is presumably associated with the volume fraction of the rocksalt phase, corresponding to the decrease in the wurtzite XRD peak intensity with higher Sc content (Figure 1(b)). Figure 3(b) maps the threshold voltage in thickness – composition space. Pure wurtzite regions where the Sc content < 0.35 do not exhibit an exponential increase in leakage current, so the threshold voltage is > 20 V for these film thicknesses. Interestingly, the thinner region (140 – 150 nm) shows relatively higher threshold voltages even for high Sc contents, which implies that the rocksalt phase fraction varies depending on the thickness. Indeed, XRD profiles of thinner regions show smaller changes in the wurtzite phase peak intensity as a function of Sc content compared to that of thicker regions (Figure 1(b)) as shown in Figure S3 (Supporting Information).

The pure wurtzite phase region (Sc < 0.35) exhibits robust ferroelectric behavior. The compositional dependence of polarization – electric field hysteresis loops taken at 10 kHz for the similar thickness points ranging from 230 – 260 nm (same y position on the sample library) is shown in Figure 3(c). Coercive field and polarization values decrease with increasing Sc content up to 0.35, in agreement with prior reports.[20,21] A device at Sc = 0.38 (green loop) breaks down at 2500 kV cm$^{-1}$, which is below the coercive field, so that no hysteresis is observed. This breakdown field relates to the large leakage current when the rocksalt phase is present. Positive-up-negative-down (PUND) measurements are carried out to quantify the switched polarization without the leakage contribution.[49,50] Figure 3(d) shows the remanent polarization (+$P_r$ - (-$P_r$)) values as a function of applied electric field for each composition of the 230 – 260 nm thickness devices. The pulse width is 100 µs, and duty cycle ratio is 10 %. The decrease in switchable polarization with increasing Sc content is consistent with the P-E hysteresis measurements.

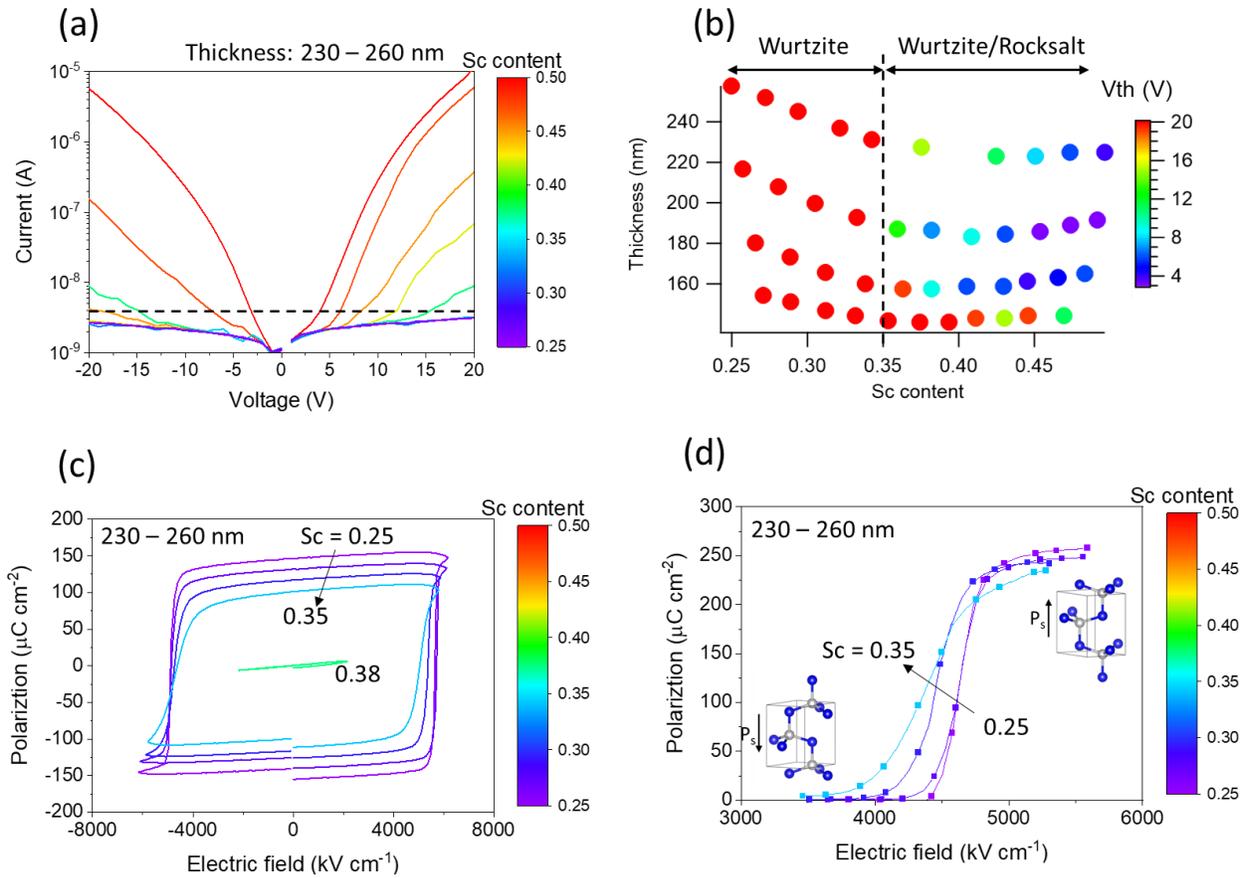

Figure 3 Electrical properties of the $Al_{1-x}Sc_xN$ library. (a) Leakage current for films with varying Sc content from 0.25 to 0.5 and film thicknesses from 230 – 260 nm. (b) Threshold voltage $V_{th}$ map in thickness – composition space showing significant conduction for Sc contents >0.35. (c) P-E hysteresis loops for the subset of films with Sc contents from 0.25 to 0.38. (d) Switched polarization under various applied electric field pulses using the PUND method for films with Sc contents from 0.25 to 0.35.

Figure 4 summarizes the ferroelectric properties, *c/a* ratio (marker size) and film thickness (marker color) as a function of Sc content in our combinatorial library compared with homogeneous films reported in systematic composition – property studies (dotted and dashed lines),[20,21] and from our chamber with nominally identical deposition condition and characterization method (black circles).[23,38] The coercive field measured at 10 kHz and remanent polarization from PUND measurements both consistently decrease with Sc content (Figure 4(a) and (b)). Note that the coercive fields of homogeneous films are results of our films using the same measurement condition, since the coercive field has a strong frequency dependence.[51] Despite the agreement in ferroelectric property change with literature, the *c/a* ratio in the library does not significantly change in the composition range as seen in the plot size in Figure 4(a) and (b). As discussed above in Figure 1(d), the *c/a* ratio is invariant with composition across the pure wurtzite region due to elastic interaction in the composition gradient library.

The key findings of this work come from comparing our combinatorial library to literature values and trends relating composition, crystallography, and ferroelectric properties. Prior studies have observed consistent and simultaneous reductions in *c/a* ratio, coercive field, and switchable polarization with increases in Sc content.[21] Crucially, the fact that Sc content is correlated with a reduction in *c/a* ratio has been repeatedly identified as the key mechanistic factor enabling ferroelectric polarization reversal in this system, but our results suggest that the underlying nature of the Sc-N chemical bonds are more important to enabling polarization reversal than simply via a change in the crystallographic *c/a* ratio. Indeed, Landau-Devonshire thermodynamic analysis suggests that the ferroelectric properties of $Al_{1-x}Sc_xN$ are less sensitive to lattice strain effects compared to classic perovskite ferroelectric $PbTiO_3$.[52] *Therefore, when learning from the $Al_{1-}$* $_x$Sc$_x$N *system to engineer other ferroelectric wurtzites, it is important to look beyond the simple lattice parameters.*

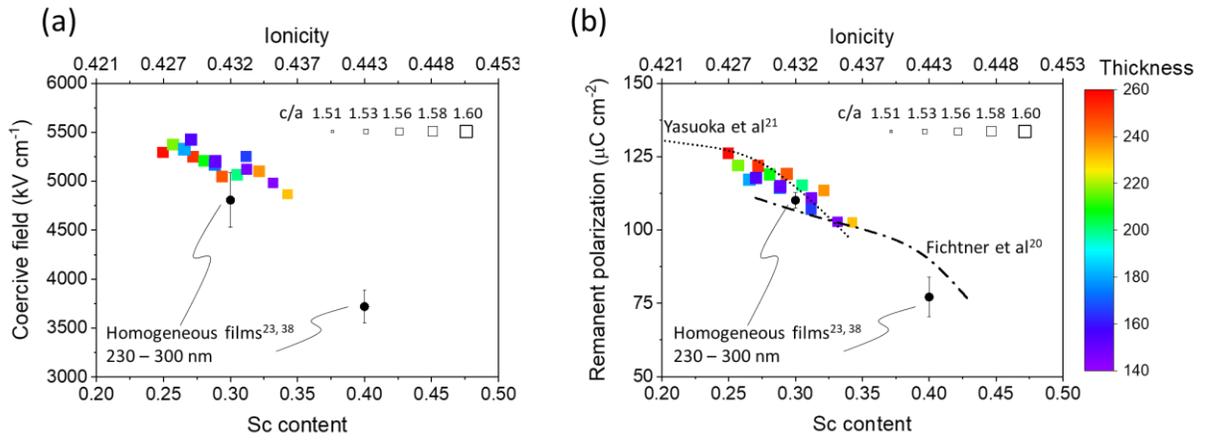

Figure 4 (a) Coercive field and (b) remanent polarization as a function of Sc content along with *c/a* and thickness in library, compared to reported homogeneous films.[20,21,23,38]

The importance of bond chemistry on ferroelectric properties has long been recognized in the oxide ferroelectrics (e.g., see frequent references to cations such as Ti, Ta, and Nb as "ferroelectrically active" and Mg, Zn, and Sc as "ferroelectrically inactive" when in octahedral coordination[53]), but such concepts require further consideration for the wurtzite nitrides. Considering classic Pauling electronegativity, the ionicity of the Al-N bond is 0.4 while the Sc-N bond is > 0.5 ionic.[54] Al-N bonding is dominated by directional $sp^3$ bonds whereas Sc-N is better described by non-directional ionic bonding, consistent with the endmember crystal structures (tetrahedral coordination in wurtzite AlN vs. octahedral coordination in rocksalt ScN). As shown on the secondary abscissa in Fig. 4(a) and (b), the average bond ionicity increases with increased Sc content. It stands to reason that the reduced directionality of such bonding would reduce barriers to polarization reversal.

We computationally modeled $Al_{1-x}Sc_xN$ alloys and performed density functional theory (DFT) calculations at 8 different compositions in the x range of 0.056-0.389 to better understand the differences in Al-N and Sc-N bonding character (see computational methods for details). Figure 5(a) illustrates the average first nearest neighbor Al-N and Sc-N bond lengths as a function of composition *x*, with error bars representing the standard deviation of the bond lengths in the supercell at a given composition. The significant difference in the Al-N and Sc-N bond lengths could be attributed to the differences in the ionic radii of Al and Sc, and to the unfavorable tetrahedral coordination of $Sc^{3+}$ (prefers octahedral coordination).

While Figure 5(a) suggests local structural differences in the neighborhood of Al and Sc cations, it does not provide insights into the nature of the chemical bonds. To investigate the cation-N bonding, we calculated the electron localization function (ELF). The calculated ELF at $x = 0.25$ is shown in Figure 5(b) as a cross-section across the c-axis. Al-N bonds, indicated by black arrows, are characterized by large ELF values along the bonds suggesting directional polar covalent bonds consistent with sp$^3$ hybridization. In contrast, ELF contours are more isotropic and centered around Sc and neighboring N atoms, which is indicative of non-directional ionic bonds. We arrive at similar conclusions about the bonding nature when examining the ELF maps at other compositions.

The calculated Born effective charges (BECs) and Bader charges provide further evidence that Sc-N bonds are more ionic than Al-N bonds. The BECs of Al and N aggregated across the 8 compositions are presented in Figure 5(c). There is a clear demarcation between Al and Sc BECs, with an average BEC of 2.62±0.07 for Al and 3.03±0.14 for Sc. BEC of Sc is closer to the formal charge of +3 while that of Al is much smaller than 3, again supporting the greater ionicity of Sc-N compared to polar covalent Al-N bonds. Our calculations also show that Sc with an average Bader charge of 1.6 is more ionic than Al (~1.25). It has been previously argued that this well-known difference in the Bader charge[55,56] and nominal charge is meaningful and reflects the degree of iconicity of the bond.[55] While both Al and Sc are cations in +3 formal oxidation states, the difference in their Bader charges (Sc 1.6 vs. Al 1.25) clearly indicates a difference in the Sc-N and Al-N bond ionicity.

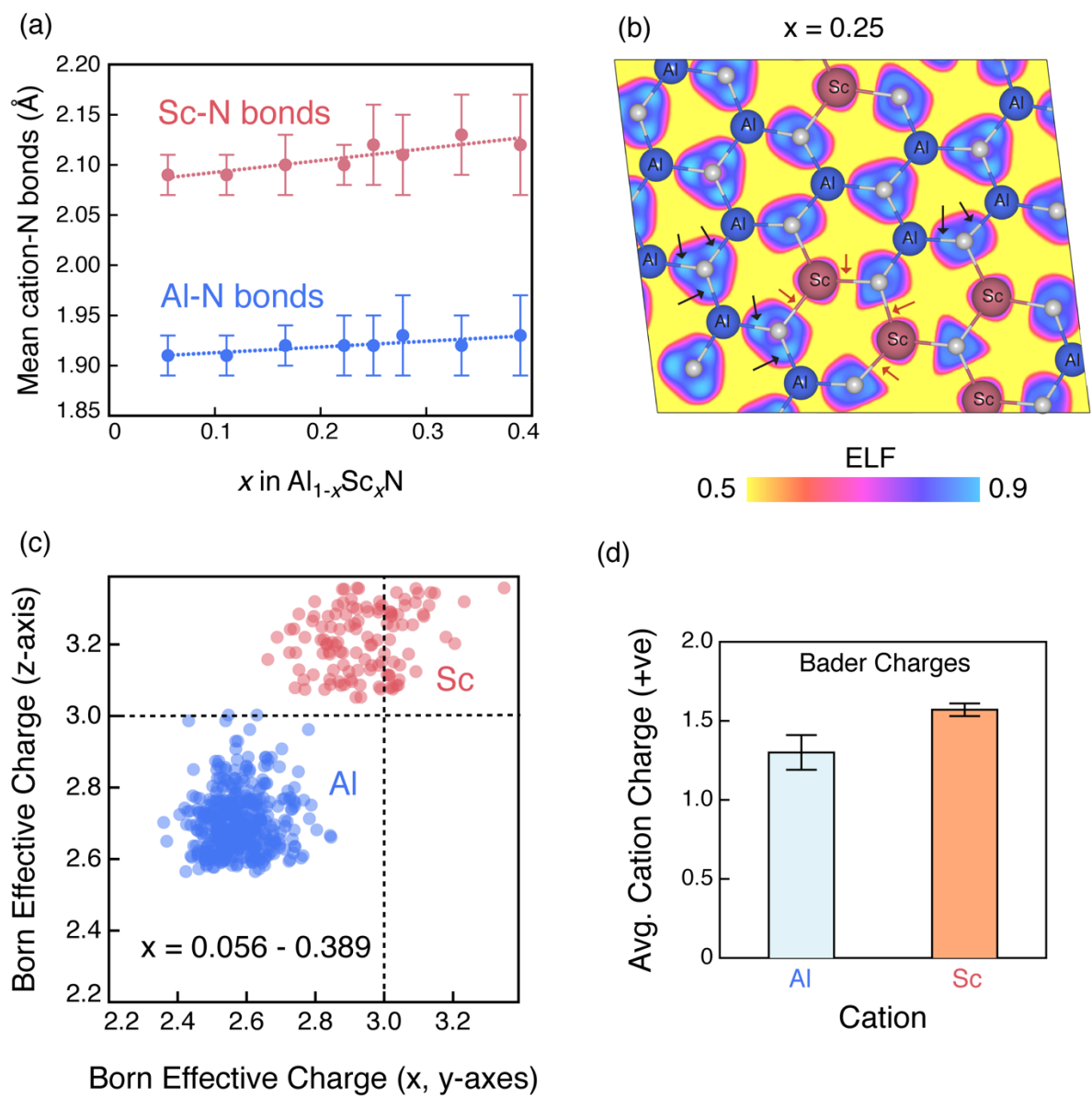

Figure 5 (a) Average first nearest neighbor Al-N and Sc-N bond lengths as a function of composition $x$ in DFT-relaxed SQS supercells. The errors bars are the standard deviations in the bond lengths. (b) Cross-sectional electron localization function (ELF) heat map computed at $x = 0.25$ composition. The black and red arrows point to representative Al-N and Sc-N bonds, respectively, showing the differences in ELF. (c) Calculated Born effective charges of Al and Sc aggregated across 8 different compositions in the range $x$ = 0.056-0.389. (d) Average Bader charges of Al and Sc (averaged across compositions). Error bars denote the standard deviation. The higher Bader charge of Sc is consistent with more ionic Sc-N bonds.

The importance of local chemical bonding as a controlling factor of ferroelectric properties proposes a guide for tetrahedral-based ferroelectric material design and discovery in the spirit of the "ferroelectrics everywhere" concept of revisiting known polar systems as potential ferroelectrics.[57] For example, the large electronegativities of oxygen and fluorine are expected to decrease coercive field making it easier to realize spontaneous polarization switching, all else being equal. Similarly, elements with low electronegativity such as alkali metal, alkaline earth, or rare-earth cations are expected to increase ionicity. Thus, oxide or halide wurtzite materials based on alkali-earth/rare-earth cations (e.g., AlN, GaN, ZnO, AgI, and CdS) can be a good candidates for ferroelectricity.[58] Indeed, $Zn_{1-x}Mg_xO$, which has a larger ionicity ranging from 0.55 to 0.68,[54] exhibits a lower coercive field (~ 3000 $kVcm^{-1}$ for x = 0.34 at 100 Hz) compared to $Al_{1-x}Sc_xN$ with ionicity ranging from 0.40 to 0.51, even with higher $c/a$ ratio (= 1.595) for $Zn_{1-x}Mg_xO$[57] compared to $Al_{1-x}Sc_xN$. As another example, $Al_{1-x}Yb_xN$ is another nitride material system that undergoes mechanical softening,[59,60] but large differences in ionic radii complicate both solubility and polarization reversal. Also, incorporation of more ionic elements into the $sp^3$ bonded covalent tetrahedral framework to ease polarization reversal has a tradeoff between of disrupting the tetrahedral network and leading to phase separation, as seen in the $Al_{1-x}Sc_xN$ system with large Sc content.[20]

In summary, tuning of crystal structure and ferroelectric properties of a combinatorial co-sputtered $Al_{1-x}Sc_xN$ film with simultaneous composition and thickness gradients is demonstrated and analyzed by theoretical calculations. Pure wurtzite phase is achieved at Sc < 0.35, whereas wurtzite and rocksalt phase coexistence is seen for Sc > 0.35 for the thickness range (140 - 260 nm). The $c/a$ ratio remains constant in the pure wurtzite region presumably due to the elastic interaction among neighboring grains. Leakage current exponentially increases in the wurtzite–rocksalt phase coexistence region, which is likely attributed to the higher conductivity of the smaller band gap rocksalt phase. Well saturated polarization – electric field hysteresis loops are observed in the pure wurtzite phase region, and coercive field and remanent polarization decrease with increasing Sc content, as previously reported. However, comparing ferroelectric properties, composition, and crystallographic $c/a$ ratio, we find that the ferroelectric response of $Al_{1-x}Sc_xN$ is controlled by the difference in chemical bonding between more ionic Sc-N bonding and covalent Al-N $sp^3$ bonding, rather than by the $c/a$ ratio as sometimes stated in prior literature. Indeed, DFT calculations reveal that Sc cations possess a more isotropic electron distribution and a BEC of 3.03±0.14 close to the formal charge, while Al cations have directional electron distribution towards nitrogens and BEC of 2.62±0.07. These results point to studying material candidates with high ionicity for decreasing coercive field of future ferroelectric wurtzite materials for memory and other applications.

**Experimental Section**

*Film Growth*: A co-sputtered $Al_{1-x}Sc_xN$ thin film with composition and thickness gradients is deposited on a $Pt/TiO_x/SiO_2/Si$ substrate using reactive RF magnetron sputtering (Figure S4(a) in Supporting Information). Elemental Al and Sc targets are located in a sputter-up configuration along the direction of the corners of the 2" $Pt/TiO_x/SiO_2/Si$ substrate, resulting in orthogonal thickness and composition gradients (Figure S4(b) in Supporting Information). Figure S4(c) and (d) in Supporting Information show the thickness and composition map of the deposited film. The color is interpolated based on an array of discrete measurement points shown as white circles. The Sc content ranges from 0.25 to 0.5 across the x direction parallel to the Al and Sc target configuration, and film thickness varies between 140 and 260 nm across y direction, perpendicular to x. Other deposition conditions are as follows: 3 mtorr of Ar/N$_2$ (13.9/4.6 sccm flow), substrate temperature of 400 °C, 90 W on a 2" diameter Al target, and 98 W on a 2" diameter Sc target. The base pressure, partial oxygen and partial water vapor pressure at 400 °C are $< 5 \times 10^{-8}$ torr, $P_{O2} < 2 \times 10^{-8}$ torr and $P_{H2O} < 1 \times 10^{-8}$ torr, respectively.

*Film Characterization*: The crystal structure of the film is investigated using X-ray diffraction (XRD) on Bruker D8 Discover and Panalytical Empyrean diffractometers. Cross-sectional transmission electron microscope (TEM) specimens were prepared by focused ion beam (FIB) in an FEI Nova NanoLab 200 dual beam FIB workstation using standard lift-out methods.[61] Final thinning of the TEM specimen was done in the FIB with a Ga+ beam energy of 5 kV and current of 41 pA. Cross-sectional TEM micrographs were acquired on a Philips CM30 TEM operated at 300 kV accelerating voltage. For electrical characterizations, top Au/Ti contacts are deposited on the film via electron beam evaporation through a shadow mask. Current-voltage (I-V) curves are measured using a bent probe whose gauge is ~300 μm contacting the bare surface of the library and a second probe contacting the exposed bottom Pt electrode. Ferroelectric properties of the Au/Al$_{1-x}$Sc$_x$N/Pt stacks were collected using a Precision Multiferroic system from Radiant Technologies. High throughput data analysis is carried out using COMBigor,[62] the files have been harvested using research data infrastructure (RDI),[63] and will be made publicly available through High Throughput Experiment Materials Database (HTEM DB).[64]

*DFT Calculation*: Special quasirandom structures (SQS)[65] were used to model the wurtzite Al$_{1-x}$Sc$_x$N alloys at 8 different compositions in the range $x = 0.056$-$0.389$. SQSs are constructed through a stochastic search over many possible configurations of local environments within the chosen supercell to best reproduce the pairwise correlation of random alloys. The SQS supercells at each composition were constructed with the Alloy Theoretic Automated Toolkit (ATAT) code.[66] Following the methodology in a previous study,[39] we constructed 72-atom SQS supercells and performed structural relaxations (volume, cell shape, and atomic positions) with density functional theory (DFT). The DFT calculations were performed with the Vienna Ab Initio Software (VASP) package.[67] The generalized gradient approximation (GGA) of Perdew-Burke-Ernzerhof (PBE)[68] was used as the exchange-correlation functional with a plane-wave energy cutoff of 340 eV. The supercells were relaxed with Γ-centered $k$-point mesh. The electron localization function (ELF) and Bader charges were calculated on a dense $k$-mesh with a fixed number of $k$-points. The $k$-point grid is determined according to the equation $n_{kpts} \times n_{atoms} \approx 8000$, where $n_{kpts}$ is the number of $k$-points and $n_{atoms}$ is the number of atoms in the supercell. The Bader charges were calculated using the code developed by Henkelman and coworkers.[69] For calculating the Born effective charges, the SQS supercells were relaxed with stricter convergence criteria; a higher plane-wave energy cutoff of 520 eV and $k$-point grid corresponding to $n_{kpts} \times n_{atoms} \approx 2000$ was used. The structures were relaxed until the energies were converged to below $10^{-8}$ eV and forces below 0.005 eV/A. Born effective charges were calculated with density functional perturbation theory, as implemented in VASP.

**Supporting Information**

Supporting Information is available from the Wiley Online Library or from the author.

**Acknowledgements**

This work was co-authored by Colorado School of Mines and the National Renewable Energy Laboratory, operated by the Alliance for Sustainable Energy, LLC, for the U.S. Department of Energy (DOE) under Contract No. DE-AC36-08GO28308. Funding was provided by the Office of Science (SC), Office of Basic Energy Sciences (BES) as part of the Early Career Award "Kinetic Synthesis of Metastable Nitrides" (material synthesis); and by the DARPA Tunable Ferroelectric Nitrides (TUFEN) program (DARPA-PA-19-04-03) as a part of Development and Exploration of FerroElectric Nitride Semiconductors (DEFENSE) project (structural and electrical characterization). The data affiliated with this study are available from the corresponding author upon reasonable request. The views expressed in the article do not necessarily represent the views of the DOE or the U.S. Government.

**Conflict of Interest**

The authors declare no conflict of interest

**Keywords**

ferorelectrics, combinatorial, nitride, wurtzite, ionicity